\documentclass[prb,twocolumn,showpacs]{revtex4}
\usepackage{amsmath}
\usepackage{graphicx}
\usepackage{times}
\usepackage{dcolumn}
\usepackage{bm}

\def\be{\begin{equation}}
\def\ee{\end{equation}}
\def\bea{\begin{eqnarray}}
\def\eea{\end{eqnarray}}
\def\bse{\begin{subequations}}
\def\ese{\end{subequations}}

\def\be{\begin{eqnarray}}
\def\ee{\end{eqnarray}}

\begin{document}

\title{Ginzburg-Landau theory for the conical cycloid state in
multiferroics: applications to CoCr$_2$O$_4$ }
\author{Chuanwei Zhang$^{1,2}$}
\author{Sumanta Tewari$^{1,3}$}
\author{John Toner$^{4}$}
\author{S. Das Sarma$^{1}$}
\affiliation{$^{1}$Condensed Matter Theory Center, Department of Physics, University of
Maryland, College Park, MD 20742\\
$^{2}$Department of Physics and Astronomy, Washington State University,
Pullman, WA 99164\\
$^{3}$Department of Physics and Astronomy, Clemson University, Clemson, SC
29634 \\
$^{4}$Department of Physics and Institute of Theoretical Science, University
of Oregon, Eugene, OR 97403}

\begin{abstract}
We show that the cycloidal magnetic order of a multiferroic can arise
in the absence of spin and lattice anisotropies, for e.g., in a cubic
material, and this explains the occurrence of such a state in CoCr$_2$O$_4$.
We discuss the case when this order 
coexists with ferromagnetism in a so called `conical cycloid' state, and
show that a direct transition to this state from the ferromagnet is
necessarily first order. On quite general grounds, the reversal of the
direction of the \textit{uniform} magnetization in this state can lead to
the reversal of the electric polarization as well, without the need to
invoke `toroidal moment' as the order parameter. 
\end{abstract}

\pacs{75.80.+q,77.80.Fm,75.30.Fv,75.10.-b}
\maketitle

\section{Introduction}

Ferromagnetism and ferroelectricity are two of the most well-known and
technologically relevant types of long range ordering that can occur in
solids. It is therefore of paramount interest and importance that in a class
of ternary oxides, known as \textquotedblleft
multiferroics\textquotedblright , both types of order seem to coexist with
the possibility of interplay between long range magnetism and long range
electric polarization \cite{Fiebig, Mostovoy1,Tokura5,Ramesh}. The recently
discovered new class of multiferroics with strong magnetoelectric effects
often display the coexistence of a spatially modulated magnetic order,
called `cycloidal' order, and uniform polarization ($\mathbf{P}$), which is
induced by the broken inversion symmetry due to the modulation of the
magnetization \cite{Mostovoy2,Lawes}. Since $\mathbf{P}$ is inherently of
magnetic origin, unusual magnetoelectric effects, as displayed by the
ability to tune the polarization by a magnetic field which acts on the
cycloidal order parameter, are possible, opening up many applications \cite%
{Mostovoy1,Tokura2,Cheong2,Chapon,Goto,Kenze,Pimenov,Sneff,Tokura3,Tokura}.
Among this exciting class of materials, the cubic spinel oxide CoCr$_{2}$O$%
_{4}$ is even more unusual, since it displays not only a non-zero $\mathbf{P}
$ and a spatially \textit{modulated} magnetic order, but also a \textit{%
uniform} magnetization \cite{Tokura} ($\mathbf{M}$) in a so-called `conical
cycloid' state (see below). The uniform component of $\mathbf{M}$ provides
an extra handle \cite{Mostovoy1} with which to tune $\mathbf{P}$, as has
been recently demonstrated \cite{Tokura}. The low value of the required
tuning magnetic field $\sim .5$ T, makes this material even more
experimentally appealing.

The ability to tune $\mathbf{P}$ by tuning the uniform part of $\mathbf{M}$
poses a theoretical puzzle, since, in existing theories, the uniform piece
of $\mathbf{M}$ should not influence the polarization at all \cite{Katsura1,
Mostovoy2, Lawes,Dagotto}. This has lead to the introduction of the
`toroidal moment', $\mathbf{T}=\mathbf{P}\times \mathbf{M}$, as the real
order parameter characterizing the conical cycloid state of CoCr$_{2}$O$_{4}$
\cite{Tokura}. In this Letter, we explain this unique phenomenon and the
other interesting aspects of the physics of the conical cycloid state by
developing a phenomenological Ginzburg-Landau (GL) theory. Additionally, the
rotationally invariant form of the theory proves that both the ordinary and
the conical cycloidal orders, with the resulting multiferroicity, are
possible even in systems \textit{without} easy plane spin and easy axis
lattice anisotropies. This is important since earlier models \cite%
{Mostovoy1, Mostovoy2, Katsura2} of the cycloidal state depend crucially on
such anisotropies. However, such anisotropic models can \textit{not} explain
the presence of the cycloidal state in cubic systems like CoCr$_{2}$O$_{4}$,
where such phases are also observed despite the fact that their cubic
symmetry forbids such easy plane and easy axis anisotropies.

CoCr$_{2}$O$_{4}$, with the lattice structure of a cubic spinel, enters into
a state with a uniform magnetization at a temperature $T_{m}=93$ K.
Microscopically, the magnetization is of ferrimagnetic origin \cite{Tokura},
and in what follows we will only consider the ferromagnetic component, $%
\mathbf{M}$, of the magnetization of a ferrimagnet. At a lower critical
temperature, $T_{c}=26$ K, the system develops a special helical modulation
of the magnetization in a plane transverse to the large uniform component.
Such a state can be described by an order parameter,
\begin{equation}
\mathbf{M}_{h}=m_{1}\hat{e}_{1}\cos (\mathbf{q}\cdot \mathbf{r})+m_{2}\hat{e}%
_{2}\sin (\mathbf{q}\cdot \mathbf{r})+m_{3}\hat{e}_{3}+h.h.,
\label{Order-Parameter-1}
\end{equation}%
where $\{\hat{e}_{i}\}$ form an orthonormal triad and $h.h.$ denotes
\textquotedblleft higher harmonics\textquotedblright\ such as terms
proportional to sines and cosines of $(2n+1)\mathbf{q}\cdot \mathbf{r}$ with
integer $n$. When the pitch vector, $\mathbf{q}$, is normal to the plane of
the rotating components, the rotating components form a conventional helix
\cite{Belitz}. For $m_{3}=0$ such a state, which we call an `ordinary helix'
state, is observed in many rare-earth metals \cite{Cooper}, e.g. MnSi \cite%
{Ishikawa, Pfleiderer}, and FeGe \cite{Lundgren}. We call a helix state with
$m_{3}\neq 0$, which is observed in some heavy rare-earth metals \cite%
{Cooper}, a `conical helix' state because the tip of the magnetization falls
on the edge of a cone. A more complicated modulation arises when $\mathbf{q}$
lies \textit{in the plane} of the rotating components. For $m_{3}=0$, we
call such a state an `ordinary cycloid' state because the profile of the
magnetization resembles the shape of a cycloid. The state with $m_{3}\neq 0$
is called a `conical cycloid' state. It is easy to see that the helical, but
not the cycloidal, modulation preserves a residual symmetry under
translations and suitable simultaneous rotations about the pitch vector.%

Since $\mathbf{M}$ and $\mathbf{P}$ respectively break time reversal and
spatial inversion symmetry, the leading $\mathbf{P}$-dependent piece in a GL
Hamiltonian density, $h_{P}$, for a centrosymmetric, time reversal invariant
system with cubic symmetry is \cite{Mostovoy2},%
\begin{equation}
h_{P}=\mathbf{P}^{2}/2\chi +\alpha \mathbf{P}\cdot \mathbf{M}\times \nabla
\times \mathbf{M},  \label{h-P}
\end{equation}%
where $\chi >0$ and $\alpha $ are coupling constants. We assume that $%
\mathbf{P}$ is a slave of $\mathbf{M}$, in the sense that a non-zero $%
\mathbf{P}$ only occurs due to the spontaneous development of a magnetic
state with a non-zero $\mathbf{M}\times \nabla \times \mathbf{M}$\textbf{,}
which then, through the linear coupling to $\mathbf{P}$ in (\ref{h-P}),
induces a non-zero $\mathbf{P}$. For an order parameter ansatz given by Eq.~%
\ref{Order-Parameter-1}, the macroscopic polarization, $\mathbf{\bar{P}}$,
is given by minimizing the Hamiltonian density (\ref{h-P}) over $\mathbf{P}$%
, $\mathbf{\bar{P}}=\chi \alpha m_{1}m_{2}[\hat{e}_{3}\times \mathbf{q}]$.
So $\mathbf{\bar{P}}$ is normal to both $\mathbf{q}$ and the axis of
rotation, $\hat{e}_{3}$. Note that in a conventional spin density wave state
($m_{1}\hspace{1.5mm}\text{or}\hspace{1.5mm}m_{2}=0$), as in the helix
states, $\mathbf{\bar{P}}$ is zero. However, for a cycloid state, $\mathbf{q}%
\perp \hat{e}_{3}$, so there is a non-zero $\mathbf{\bar{P}}$. Note that $%
\mathbf{\bar{P}}$ is entirely due to the cycloidal components $m_{1}$ and $%
m_{2}$, and is independent of the uniform magnetization $m_{3}$. Thus, while
it is conceivable that magnetic fields strong enough to `flop' the spins and
the axis of rotation of the cycloidal components will alter $\mathbf{\bar{P}}
$ \cite{Mostovoy2,Lawes,Tokura2,Cheong2}, no explanation of how tuning the
uniform component of $\mathbf{M}$ can affect the induced polarization has
been offered. We will do so later in this paper.

The paper is organized as follows: Section II lays out the Ginzburg-Landau
Hamiltonian and the parameter regions which exhibits the cycloidal phase.
Section III and V are devoted to the phase diagrams of ordinary cycloidal
state and conical cycloidal state respectively. In Section V, we explain why
the reversal of the direction of the uniform magnetization in the conical
cycloidal state can lead to the reversal of electric polarization. Section
VI consists of conclusions.

\section{Ginzburg-Landau Hamiltonian}

We consider a Hamiltonian that is \textit{completely} invariant under
simultaneous rotations of positions and magnetization. This guarantees that
any phase that can occur in our model is \textit{necessarily} allowed in a
crystal of \textit{any} symmetry. The full Hamiltonian is given by, $H=\int
(h_{M}+h_{P})d\mathbf{r}\equiv \int hd\mathbf{r}$. Using $\mathbf{P=-}\chi
\alpha \mathbf{M}\times \nabla \times \mathbf{M}$ to eliminate $\mathbf{P}$,
we can write the total Hamiltonian density $h$ entirely in terms of $\mathbf{%
M}$,
\begin{eqnarray}
h &=&t\mathbf{M}^{2}+u\mathbf{M}^{4}+K_{0}\left( \nabla \cdot \mathbf{M}%
\right) ^{2}+K_{1}\left( \nabla \times \mathbf{M}\right) ^{2}  \notag \\
&&+K_{2}\mathbf{M}^{2}\left( \nabla \cdot \mathbf{M}\right) ^{2}+K_{3}\left(
\mathbf{M}\cdot \nabla \times \mathbf{M}\right) ^{2}  \notag \\
&&+K_{4}\left\vert \mathbf{M}\times \nabla \times \mathbf{M}\right\vert ^{2}
\notag \\
&&+D_{L}|\nabla \left( \nabla \cdot \mathbf{M}\right) \mathbf{|}%
^{2}+D_{T}|\nabla \left( \nabla \times \mathbf{M}\right) \mathbf{|}^{2},
\label{h-M}
\end{eqnarray}%
where we have $u$, $D_{L,T}>0$ for stability. In Eq.~\ref{h-M}, where the
Landau expansion of the free energy is truncated at the fourth order, the
usual gradient-squared term, $c\left\vert \nabla \mathbf{M}\right\vert ^{2}$%
, is omitted since, $\left\vert \nabla \mathbf{M}\right\vert ^{2}=(\nabla
\cdot \mathbf{M})^{2}+|\nabla \times \mathbf{M}|^{2} $, plus an unimportant
surface term which can be neglected. Notice that, for $K_{0}=K_{1}$ and $%
K_{2}=K_{3}=K_{4}$, $h$ is rotationally invariant in the spin space alone,
so the $K_{i}$'s themselves are not proportional to the spin-orbit coupling
constant (for e.g., via the above identity, $K_{0},K_{1}\sim c$). However,
the \textit{difference} among the $K_{i}$'s should be small due to the
smallness of the spin-orbit coupling. The effects of the competing magnetic
interactions, which are present in the multiferroics and are responsible for
the spatial modulation of $\mathbf{M}$ \cite{Mostovoy1, Mostovoy2, Dagotto,
Katsura2}, are embodied in $K_{0},K_{1}$, which can be negative leading to a
spatially modulated order parameter. For decoupled spin and coordinate
spaces ($K_{i}$'s equal), the energies of the helical and the cycloidal
modulations of the spins are identical. In a system where the spin
anisotropy constrains the spins to lie on a plane, and the lattice
anisotropy forces $\mathbf{q}$ to be also on that plane, the energy of the
cycloidal modulation can be lower than that of the helical modulation \cite%
{Mostovoy2, Katsura2}. Such anisotropies have been implicitly taken as the
driving force behind the cycloidal order by Mostovoy \cite{Mostovoy2}, and
Katsura \textit{et al.} \cite{Katsura2}. For cubic crystals, however, no
such anisotropy exists among the principal directions. We argue below that,
in this case, the magnetoelectric couplings themselves, leading to the
difference among the $K_{i}$'s, can lower the energy of the cycloidal state
than that of any other state with an arbitrary angle between $\mathbf{q}$
and the plane of the magnetization. 

Rather than exploring the complete parameter space of this model, we limit
ourselves to two different parameter regions, which exhibit all the phases
described above:

\textbf{Region I}: $K_{0,1}<0,K_{i>1}$ small, $t>0$ , and

\textbf{Region II}: $t<0$, $K_{3}<0$, $K_{1}>0$, $K_{2}=K_{4}=0$.

We have checked that our results are robust against allowing small non-zero
values of the various $K_{i}$'s that we take to be zero. In that sense our
results, in particular the topology of the phase diagrams shown in Figs. \ref%
{Figure-Phase-Diagram1}a and \ref{Figure-Phase-Diagram2}a for Regions I and
II, respectively, and the orders of the various phase transitions that we
predict, are generic. As usual, our theoretical phase diagrams can be
related to experimental ones by noting that \textit{all} of the
phenomenological parameters $(t,K_{i},D_{L,T},u)$ in our model should depend
on experimental parameters like, e.g., temperature ($T$). Thus, an
experiment in which, e.g., $T$ is varied with all other parameters held
fixed will map out a locus of points through our theoretical phase diagrams.
In Landau theories, $t$ is expected to vary from large positive values,
corresponding to disordered phases with $\mathbf{M}(\mathbf{r})=\mathbf{0}$,
at high $T$, to smaller values at which $\mathbf{M}(\mathbf{r})\neq \mathbf{0%
}$ become possible. In order to access the conical cycloid state, we must
also allow $K_{0}(T)$ and $K_{1}(T)$ to change sign as $T$ is decreased.

For the most part we will work in mean field theory, which is simply finding
a magnetization configuration $\mathbf{M}(\mathbf{r})$ that minimizes the
Hamiltonian (\ref{h-M}). Clearly, the task of finding the \textit{global}
minimum is a formidable one. Instead, we restrict ourselves to ansatzes of
the form:
\begin{equation}
\mathbf{M}=m_{1}\hat{e}_{1}\cos (\mathbf{q}\cdot \mathbf{r})+m_{2}\hat{e}%
_{2}\sin (\mathbf{q}\cdot \mathbf{r})+\mathbf{M}_{0},
\label{Order-Parameter-ansatz}
\end{equation}%
where the spatially constant vector $\mathbf{M}_{0}$ is allowed to point in
\textit{any} direction. (Given the global rotation invariance under
simultaneous rotations of magnetization and space, an infinity of other
solutions trivially related to (\ref{Order-Parameter-ansatz}) by such
rotations, and with exactly the same energy, also exist, of course.) In the
special case of $\mathbf{q}$ along $x$ direction (or, equivalently, anywhere
in the $x-y$-plane), this is a cycloid state with a uniform background
magnetization $\mathbf{M}_{0}=\left( M_{01},M_{02},M_{03}\right) $. When $%
\mathbf{q}$ is along $z$ direction, it is a helix state. Inserting this
ansatz (\ref{Order-Parameter-ansatz}) into the Hamiltonian (\ref{h-M}), and
integrating over the volume $V$ of the system, we can obtain the energy of
the system. Through the minimization of the energy, we find the conical
cycloid state is the \textit{only} state with a non-zero $\mathbf{M}_{0}$
when $K_{3}<K_{4}$. In addition, the optimal direction for $\mathbf{q}$ is
\textit{always} either in the ($x-y$) plane, or orthogonal to it. Putting
these facts together means that \textit{all} of the minimum energy
configurations are of the form ({\ref{Order-Parameter-1}). Furthermore, when
$\mathbf{q}$ lies in the ($x-y$) plane, we can always use the global
rotation invariance of our model to rotate $\mathbf{q}$ to lie along the $x$%
-axis, and will henceforth do so. }

\section{Ordinary Cycloid State}

\begin{figure}[b]
\includegraphics[scale=0.45]{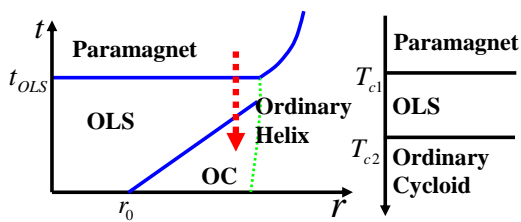}
\caption{(Color online) (a) Phase diagram in Region I for the ordinary
cycloid state. Solid (blue) lines represent second order phase transitions.
Dotted (green) line indicates the first order transition to the helix state.
The dotted (red) arrow represents one possible schematic locus of the
experimental points obtained by varying $T$. $r_{0}\equiv \left(
1+3D_{T}/D_{L}\right) /6$. (b) The sequence of phases with decreasing $T$
along the locus shown.}
\label{Figure-Phase-Diagram1}
\end{figure}

In \textbf{Region I}, the dominant terms in the Hamiltonian involving the
uniform component $m_{3}$ are $tm_{3}^{2}+um_{3}^{4}$, therefore the lowest
energy states have $m_{3}=0$. Small negative $K_{i>1}$ clearly cannot change
this fact. The energy for the ordinary cycloid (OC) state is obtained by
inserting (\ref{Order-Parameter-1}) with $m_{3}=0$ into the Hamiltonian
\begin{equation}
{E/V}=\Gamma _{L}\left( q\right) m_{1}^{2}+\Gamma _{T}\left( q\right)
m_{2}^{2}+u\Phi (m_{1}^{2},m_{2}^{2}),  \label{EOC}
\end{equation}%
where $\Gamma _{L}\left( q\right) =\left( t+K_{0}q^{2}+D_{L}q^{4}\right) /2$%
, $\Gamma _{T}\left( q\right) =\left( t+K_{1}q^{2}+D_{T}q^{4}\right) /2$,
and $\Phi (m_{1}^{2},m_{2}^{2})=3\left( m_{1}^{4}+m_{2}^{4}\right)
/8+m_{1}^{2}m_{2}^{2}/4$. In writing this, we have neglected the higher
harmonics in Eq. (\ref{Order-Parameter-1}), whose amplitude vanishes much
faster (specifically, as fast or faster than $|m_{i}|^{3}$) than the
magnitude of the order parameter itself, and thus have negligible effects on
the phase boundaries. For large positive $t$, all the terms in this energy
are positive, and, hence, the lowest energy state is $m_{1}=m_{2}=0$; i.e.,
the paramagnet. As $T$ decreases, $t$ becomes smaller and the first phase
transition that will occur depends on whether the minimum over $q$ of $%
\Gamma _{L}\left( q\right) $ or $\Gamma _{T}\left( q\right) $ becomes
negative first. For $r\equiv K_{1}/K_{0}<\sqrt{D_{L}/D_{T}}$, $\Gamma
_{L}\left( q\right) $ becomes negative first at $t_{OLS}=K_{0}^{2}/4D_{L}$,
and $m_{1}$ starts to be nonzero. This boundary between paramagnet and the
ordinary longitudinal spin density wave (OLS) phase ($m_{2}=m_{3}=0$, $%
m_{1}\neq 0$) is the horizontal (solid blue ) line in the phase diagram Fig. %
\ref{Figure-Phase-Diagram1}a in the ($r,t$) plane for fixed negative $K_{0}$
and all $K_{i>1}=0$.

The OLS phase will, as we continue lowering $t$, eventually become unstable
to a non-zero $m_{2}$; this is the OC state. By minimizing the energy (\ref%
{EOC}) in the OLS phase, we find $q^{2}=q_{L,min}^{2}=-K_{0}/2D_{L}$ and $%
m_{1}^{2}={2}(t_{OLS}-t)/3u$. Inserting these into ({\ref{EOC}) we find that
the coefficient of }$m_{2}^{2}$ becomes negative below {$t_{LOC}=t_{OLS}%
\left[ 3r-\left( 1+3D_{T}/D_{L}\right) /2\right] $. This value $t_{LOC}$ of }%
$t$ therefore defines the locus of a continuous {OLS-OC phase transition,
and is the non-horizontal straight (solid (blue)) line in the $r-t$ plane
shown in Fig. \ref{Figure-Phase-Diagram1}a.}

For $r>\sqrt{{\frac{D_{L}}{D_{T}}}}$, $\Gamma _{T}$ becomes non-zero first,
which seems to imply that one enters the ordinary transverse spin density
wave (OTS) phase ($m_{1}=m_{3}=0$, $m_{2}\neq 0$) first for large $r$.
However, it is not true because the OTS phase always has higher energy than
the ordinary helical (OH) phase. The energy for the ordinary helix state is%
\begin{equation}
E/V=\Gamma _{T}(m_{1}^{2}+m_{2}^{2})+\Phi (m_{1}^{2},m_{2}^{2}).
\label{Ene-helix}
\end{equation}%
The minimization of the energy over the direction of $\left(
m_{1},m_{2}\right) $ vector yields $|m_{1}|=|m_{2}|=m_{H}/\sqrt{2}$, that
is, a \textit{circular} helix. Further minimization over $m_{H}$ and $q$
gives the energy $E_{OH}$ of the ordinary helix state $%
E_{OH}/V=-(t_{OH}-t)^{2}/4u{\ }$for $t<t_{OH}$, where $%
t_{OH}=r^{2}D_{L}t_{OLS}/D_{T}$. The energy for the OTS state is $%
E_{OTS}/V=-(t_{OH}-t)^{2}/6u$, which is obtained from equation (\ref{EOC})
by setting $m_{1}=0$ and $q^{2}=q_{T,min}^{2}\equiv -\frac{K_{1}}{2D_{T}}$,
and then minimizing over $m_{2}$. $E_{OTS}$ is clearly higher than $E_{OH}$.
Hence, the helical state is always favored over the OTS state throughout
Region I of the phase diagram. Note that $t_{OH}$ defines the boundary for
the second order transition from the paramagnet to the OH state.

There is also a direct first order phase transition between the OH and the
OLS states along the line where $E_{OH}=E_{OLS}$. Here $%
E_{OLS}/V=-(t_{OLS}-t)^{2}/6u$ is the energy for OLS state obtained from
equation (\ref{EOC}). This equality yields the first order phase boundary $%
t_{OLH}=(\sqrt{3/2}t_{OH}-t_{OLS})/(\sqrt{3/2}-1)$ between the OH and the
OLS states (the dotted (green) line). {The line for the OLS-OC transition
always intersects the first order OLS-OH phase boundary before crossing the
paramagnet-OLS boundary. This therefore always yields the topology shown in
Fig. \ref{Figure-Phase-Diagram1}a. }

A typical experimental locus through this phase diagram, namely one in which
$t$ decreases as temperature $T$ does, with $r$ constant, is shown in Fig. %
\ref{Figure-Phase-Diagram1}a. The sequence of phases that results is
illustrated in Fig. \ref{Figure-Phase-Diagram1}b. We see that the paramagnet
to ordinary cycloid phase transition is always preempted by a paramagnet to
OLS phase transition, and the cycloid state is always elliptical. Both of
these predictions are borne out by recent experiments on TbMnO$_{3}$ \cite%
{Sneff,Tokura3}. On the other hand, a direct transition to the circular
helix state is predicted by our theory, and has indeed been observed
experimentally \cite{Ishikawa, Pfleiderer}.

All of the above statements are based on mean field theory, that is theory
without considering the fluctuations. Going beyond mean field theory, very
general arguments due to Brazovskii \cite{Braz} imply that, in rotation
invariant models, \textit{any} direct transition from a homogeneous state
(paramagnet) to a translationally ordered one (OLS and OH) \textit{must} be
driven first order by fluctuations. Consideration of topological defects and
orientational order \cite{Toner,Toner2,Toner3} supports this conclusion, but
raises the additional possibility that direct transition between the
homogeneous and the translationally ordered phases could split into two,
with an intermediate orientationally ordered phase, analogous to the 2D
\textquotedblleft hexatic\textquotedblright\ phase \cite{hexatic}. In the
present context, this implies that both the paramagnet to OLS and OH phase
transitions are either driven first order by fluctuations, or split into two
transitions with an intermediate orientationally ordered phase. Crystal
symmetry breaking fields neglected in our model could invalidate this
conclusion, if strong enough.

\section{Conical Cycloid State}

\begin{figure}[t]
\includegraphics[scale=0.45]{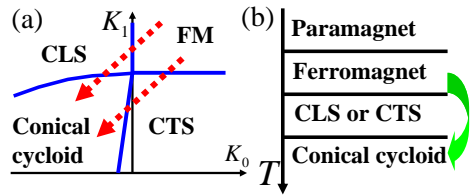}
\caption{(Color online) (a) Phase diagram in Region II for the conical
cycloid state. Solid (blue) lines are the boundaries between different
phases. The dotted (red) arrows represent possible paths for transition to
the CC state via continuous transitions. (b) The succession of the phases
with decreasing $T$. The (green) arrow represents a direct first order
transition between the FM and the CC state.}
\label{Figure-Phase-Diagram2}
\end{figure}

In \textbf{Region II}, we can show that conical cycloid (CC) state of the
form $\mathbf{M}=\left( m_{1}\cos (qx),m_{2}\sin (qx),m_{3}\right) $ is the
lowest energy state among all the possible states with arbitrary mutual
angles between the uniform magnetization, $\mathbf{q}$, and the cycloid
plane. The energy $E$ for this state takes the form%
\begin{eqnarray}
E/V &=&\left( t+K_{0}q^{2}+D_{L}q^{4}+2um_{3}^{2}\right) m_{1}^{2}/2
\label{ECC} \\
&&+\left( t+K_{1}q^{2}+D_{T}q^{4}+2um_{3}^{2}+K_{3}q^{2}m_{3}^{2}\right)
m_{2}^{2}/2  \notag \\
&&+u\Phi (m_{1}^{2},m_{2}^{2})+tm_{3}^{2}+um_{3}^{4},  \notag
\end{eqnarray}%
where we have again neglected the higher harmonics in Eq. (\ref%
{Order-Parameter-1}). In this region, the \textit{h.h.} terms do not vanish
as the conical longitudinal spin density wave (CLS) ($m_{2}=0$, $m_{1,3}\neq
0$) or conical transverse spin density wave (CTS) ($m_{1}=0$, $m_{2,3}\neq 0$%
) to FM transition in Fig. {\ref{Figure-Phase-Diagram2}} is approached.
However, we have verified that amplitudes of the \textit{h.h.} terms are
only a very small fraction of the cycloidal components $m_{1}$ and $m_{2}$
(not of the uniform component $m_{3}$), therefore their neglect below (but
close to) the lower cycloidal transition temperature of 26 K is justified.
They have little or no quantitative effect on our phase diagram or the
orders of the transition.

Since $t<0$, we can minimize Eq. (\ref{ECC}) over $m_{3}$ with $m_{1}=m_{2}=0
$, and find a ferromagnetic (FM) state with $m_{3}=\sqrt{-t/2u}$. For large
positive $K_{0}$ and $K_{1}$, this ferromagnetic state is clearly stable
against the development of non-zero $m_{1}$ and $m_{2}$. It also clearly
becomes \textit{unstable} against the development of a non-zero $m_{1}$ if $%
K_{0}$ is lowered to negative values, because then the coefficient $%
(K_{0}q^{2}+D_{L}q^{4})$ of $m_{1}^{2}$ becomes negative for sufficiently
small $q$. This instability (which is clearly into the CLS state) will occur
at $K_{0}=0$, at a wavevector $q$ satisfying $q_{L,min}^{2}=-K_{0}/2D_{L}$.
Note, however, that now, because $K_{0}$ is being varied \textit{through}
zero, this wavevector will now \textit{vanish} as the transition is
approached from below. The order parameter $m_{1}^{2}=K_{0}^{2}/2uD_{L}$
\textit{also} vanishes as this transition is approached. Thus, this
transition is, like the $\beta $ - incommensurate transition in quartz and
berlinite \cite{Biham}, simultaneously a \textit{nucleation} transition ($q$
vanishes), and an \textit{instability} transition (order parameter
vanishes). Indeed, this transition and the FM$\rightarrow $ CTS transition,
which is of the same type and will be discussed below, are, to our
knowledge, the \textit{first} examples of transitions that exhibit such a
dual character in a model \textit{without} terms linear in the gradient
operator.

We can find the loci of instability between the CLS phase and the CC state
by calculating the coefficient of $m_{2}^{2}$ in (\ref{ECC}) in the CLS
phase, and finding where it becomes negative. The minimization of the energy
(\ref{ECC}) over $q$, $m_{3}$ and $m_{1}$ yields $q^{2}=-K_{0}/2D_{L}$, $%
m_{3}^{2}=-\left( t+K_{0}^{2}/2D_{L}\right) /2u$ and $%
m_{1}^{2}=K_{0}^{2}/2uD_{L}$. Inserting these expressions into (\ref{ECC})
and taking the coefficient of $m_{2}^{2}$ to be zero, we find the CLS to CC
phase boundary as:%
\begin{equation}
K_{1}=\frac{K_{0}}{2}\left( \frac{D_{T}}{D_{L}}-1+\frac{K_{0}K_{3}}{2uD_{L}}%
\right) +\frac{tK_{3}}{2u}.  \label{clcc}
\end{equation}

Similar analysis of the sequence of the phase transition, FM $\rightarrow $
CTS $\rightarrow $ CC, yields the schematic phase diagram on the $%
K_{0}-K_{1} $ plane given in Fig. \ref{Figure-Phase-Diagram2}a. The phase
boundary between FM and CTS is given by $K_{1}=tK_{3}/2u$. The phase
boundary between the CTS and the CC phase at small $K_{0}$ is $%
K_{1}=2K_{0}/\left( D_{L}/D_{T}-1\right) $, which is also shown in Fig. \ref%
{Figure-Phase-Diagram2}a.

Fig. \ref{Figure-Phase-Diagram2} shows that it is not possible to go from
the FM to the CC state via a continuous transition, except at a single
special point. Generic paths like the diagonal dashed lines in Fig. \ref%
{Figure-Phase-Diagram2}a \textit{must} go through either the CLS or the CTS
state, so two transitions are required to reach the CC state, which,
additionally, must be elliptical. Hence the only way there can be a direct
transition from the FM state to the CC state is via a first order phase
transition, which is not addressed by our theory. This prediction is borne
out by experiments of CoCr$_{2}$O$_{4}$, where the direct FM to CC
transition is indeed first order \cite{Tokura}.

\section{Magnetic Reversal of the electric polarization:}

The polarization $\bar{\mathbf{P}}=\chi \alpha m_{1}m_{2}\hat{y}$ in the CC
state is in the $xy$ plane, normal to $\hat{e}_{3}$ and $\mathbf{q}$. It is
independent of the uniform magnetization, $m_{3}$. Experimentally \cite%
{Tokura}, the sample is cooled through $T_{c}$ in the presence of a small
electric field, $\mathbf{E}=E_{0}\hat{y}$, and a small magnetic field, $%
\mathbf{H}=H_{0}\hat{z}$. The direction of the pitch vector, $\hat{x}$, or,
equivalently, the axis of rotation, $\hat{z}$, are set by the direction of $%
\bar{\mathbf{P}}$ ($\mathbf{E}$), which determines the `helicity' of the
cycloid \cite{Tokura2}. It is found, at first, that $\bar{\mathbf{P}}$ is
uniquely determined by $\mathbf{E}$ alone, independent of the \textit{initial%
} direction of $\mathbf{H}$, as expected. However, once $\bar{\mathbf{P}}$
and $m_{3}$ have set in, changing $H_{0}$ to $-H_{0}$ not only reverses the
direction of $m_{3}$, but also, quite unexpectedly, reverses the direction
of $\bar{\mathbf{P}}$ as well. In the literature \cite{Mostovoy1, Tokura},
this has lead to the definition of the `toroidal moment', $\mathbf{T}=%
\mathbf{P}\times \mathbf{M}$, 
as the order parameter.

It is clear that the experimental system is in the conical cycloid state,
where $m_{3},\mathbf{q}$ and $\bar{\mathbf{P}}$ are always in mutually
orthogonal directions \cite{Tokura}. Further, as expected for this state,
the directions of $m_{3}$ and $\bar{\mathbf{P}}$ are uniquely determined by
the small cooling fields, $\mathbf{H}$ and $\mathbf{E}$, respectively, which
add terms to the Hamiltonian that split the degeneracy between the minima
corresponding to the different directions. Now assume that the direction of $%
\mathbf{H}$ is reversed, $H_{0}\rightarrow -H_{0}$, reversing the direction
of $m_{3}$ once it has well developed. There are two ways the uniform
magnetization can reverse its direction. First, $m_{3}$ may continue to
remain along the $z$-axis and its magnitude may pass through zero to become $%
-m_{3}$ for $\mathbf{H}=-H_{0}\hat{z}$. If this is the case, $\bar{\mathbf{P}%
}$ will remain fixed in the direction $\hat{y}$, since the mutual
orthogonality of $m_{3},\mathbf{q}$ and $\bar{\mathbf{P}}$ can always be
maintained and there is no direct coupling between $m_{3}$ and $\bar{\mathbf{%
P}}$. However, since $m_{3}$ is already well developed and large ($T_{m}=93$
K), due to the magnetic exchange energy cost it may be energetically more
favorable to leave the magnitude of $m_{3}$ unchanged, and its direction may
\textit{rotate in space} to $-\hat{z}$. If this is the case, then $m_{3}$
must rotate staying on the $y-z$ plane, since that way it always remains
perpendicular to $\mathbf{q}$, whose direction fluctuations cost the
crystalline anisotropy energy. It is then clear, see Fig.~\ref%
{Figure-Rotation}, that the cycloid plane itself, which is always
perpendicular to $m_{3}$ to maintain the lowest energy configuration, must
rotate about $\hat{x}$ by a total angle $\pi $. It follows that $\bar{%
\mathbf{P}}$, always on the cycloid plane, reverses its direction to $-\hat{y%
}$. This way, even though there is no dynamical coupling between $m_{3}$ and
$\bar{\mathbf{P}}$, the latter can also \textit{rotate} by an angle $\pi $
as a result of the former reversing its direction in space. Based on this,
we predict that, at some intermediate $\mathbf{H}\sim -H^{\prime }\hat{z}$,
where $H^{\prime }<H_{0}$, $\bar{\mathbf{P}}$ points in the direction $-\hat{%
z}$, which can be experimentally tested. 
\begin{figure}[t]
\includegraphics[scale=0.65]{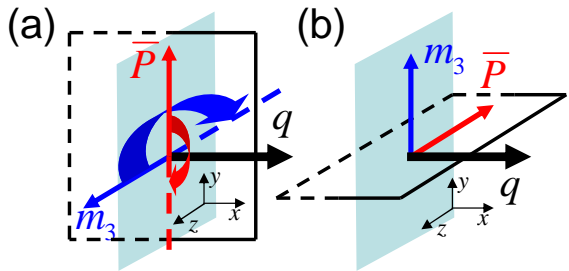}
\caption{The reversal of the polarization ($\bar{\mathbf{P}}$) by the
reversal of the magnetization ($m_{3}$). (a) If $m_{3}$ \textit{rotates} to $%
-m_{3}$, remaining perpendicular to $\mathbf{q}$, the cycloidal ($xy$) plane
must rotate accordingly to always remain transverse to $m_{3}$, which is the
lowest energy configuration. Since $\bar{\mathbf{P}}$ is in the cycloidal
plane, it will rotate by a total angle $\protect\pi $. (b) An intermediate
stage when $m_{3}$ has rotated by an angle $\frac{\protect\pi }{2}$ and
points in the $\hat{y}$ direction. At this stage, $\bar{\mathbf{P}}$ points
in the $-\hat{z}$ direction.}
\label{Figure-Rotation}
\end{figure}

\section{Conclusions}

To conclude, we've shown that the magnetic cycloidal orders, and the
resulting multiferroicity, can naturally arise due to the magnetoelectric
couplings even in rotationally invariant systems, or in cubic crystals. This
explains such orders in CoCr$_{2}$O$_{4}$, which lack easy plane
anisotropies, and are hence outside the realm of the previous theoretical
studies on multiferroics. We also predict that a second order transition
from the ferromagnet to the conical cycloid state can only occur through an
intervening conical longitudinal or transverse spin density wave state with
the ultimate cycloidal state being elliptical. A direct such transition,
then, must be first order. An important feature of our Ginzburg-Landau
theory is that we do not need to invoke an arbitrary (and ad hoc) `toroidal
moment' to explain the interplay between the magnetization and the
polarization -- the behavior which has been attributed to the toroidal
moment arises naturally in our theory.

We thank D. Drew, D. Belitz, and R.Valdes Aguilar for useful discussions.
This work is supported by NSF, NRI, LPS-NSA, and SWAN.


\begin{thebibliography}{99}
\bibitem{Fiebig} M. Fiebig, J. Phys. D: Appl. Phys. \textbf{38}, R123 (2005).

\bibitem{Mostovoy1} S.-W. Cheong and M. Mostovoy, Nature Materials \textbf{6}%
, 13 (2007).

\bibitem{Tokura5} Y. Tokura, Science \textbf{312}, 1481 (2006).

\bibitem{Ramesh} R. Ramesh and N.A. Spaldin, Nature Materials \textbf{6}, 21
(2007).

\bibitem{Mostovoy2} M. Mostovoy, Phys. Rev. Lett. \textbf{96}, 067601 (2006).

\bibitem{Lawes} G. Lawes, A. B. Harris, T. Kimura, N. Rogado, R. J. Cava, A.
Aharony, O. Entin-Wohlman, T. Yildirim, M. Kenzelmann, C. Broholm, and A. P.
Ramirez, Phys. Rev. Lett. \textbf{95}, 087205 (2005).

\bibitem{Tokura2} T. Kimura, T. Goto, H. Shintani, K. Ishizaka, T. Arima and
Y. Tokura, Nature \textbf{426}, 55 (2003).

\bibitem{Cheong2} N. Hur, S. Park, P. A. Sharma, J. S. Ahn, S. Guha, S.W.
Cheong, Nature \textbf{429}, 392 (2004).

\bibitem{Chapon} L.C. Chapon, G. R. Blake, M. J. Gutmann, S. Park, N. Hur,
P. G. Radaelli, and S-W. Cheong, Phys. Rev. Lett. \textbf{93}, 177402 (2004).

\bibitem{Goto} T. Goto, T. Kimura, G. Lawes, A. P. Ramirez, and Y. Tokura,
Phys. Rev. Lett. \textbf{92}, 257201 (2004).

\bibitem{Kenze} M. Kenzelmann, A. B. Harris, S. Jonas, C. Broholm, J.
Schefer, S. B. Kim, C. L. Zhang, S.-W. Cheong, O. P. Vajk, and J. W. Lynn,
Phys. Rev. Lett. \textbf{95}, 087206 (2005).

\bibitem{Pimenov} A. Pimenov, A. A. Mukhin, V. Yu. Ivanov, V. D. Travkin, A.
M. Balbashov, A. Loidl, Nature Phys. \textbf{2}, 97 (2006).

\bibitem{Sneff} D. Senff, P. Link, K. Hradil, A. Hiess, L. P. Regnault, Y.
Sidis, N. Aliouane, D. N. Argyriou, and M. Braden, Phys. Rev. Lett. \textbf{%
98}, 137206 (2007).

\bibitem{Tokura3} Y. Yamasaki, H. Sagayama, T. Goto, M. Matsuura, K. Hirota,
T. Arima, and Y. Tokura, Phys. Rev. Lett. \textbf{98}, 147204 (2007).

\bibitem{Tokura} Y. Yamasaki, S. Miyasaka, Y. Kaneko, J.-P. He, T. Arima,
and Y. Tokura, Phys. Rev. Lett. \textbf{96}, 207204 (2006).

\bibitem{Katsura1} H. Katsura, N. Nagaosa, and A. V. Balatsky, Phys. Rev.
Lett. \textbf{95} 057205 (2005).

\bibitem{Dagotto} I. A. Sergienko and E. Dagotto, Phys Rev. B \textbf{73},
094434 (2006).

\bibitem{Katsura2} H. Katsura, A. V. Balatsky, and N. Nagaosa, Phys. Rev.
Lett. \textbf{98}, 027203 (2007).

\bibitem{Belitz} D. Belitz, T. R. Kirkpatrick, and A. Rosch, Phys. Rev. B
\textbf{73}, 054431 (2006).

\bibitem{Cooper} B. R. Cooper, in \textit{Solid State Physics}, edited by F.
Seitz \textit{et al.} (Academic Press, NY, 1968), Vol. 21, p.293.

\bibitem{Ishikawa} Y. Ishikawa, K. Tajima, D. Bloch, M. Roth, Solid State
Commun. \textbf{19}, 525 (1976).

\bibitem{Pfleiderer} C. Pfleiderer, G. J. McMullan, S. R. Julian, and G. G.
Lonzarich, Phys. Rev. B \textbf{55}, 8330 (1997).

\bibitem{Lundgren} L. Lundgren, O. Beckman, V. Attia, S. P. Bhattacheriee,
and M Richardson, Phys. Scr. \textbf{1}, 69 (1970).

\bibitem{Braz} S. A. Brazovskii and S. G. Dmitriev, JETP \textbf{42}, 497
(1976).

\bibitem{Toner} D. R. Nelson and J. Toner, Phys. Rev. B \textbf{24}, 363
(1981).

\bibitem{Toner2} J.\ Toner, Phys.\ Rev.\ A \textbf{27}, 1157 (1983).

\bibitem{Toner3} G.\ Grinstein, T.\ C.\ Lubensky, and J.\ Toner, Phys.\
Rev.\ B \textbf{33}, 3306 (1986).

\bibitem{hexatic} B. I. Halperin and D. R. Nelson, Phys. Rev. Lett. \textbf{%
41}, 121 (1978).

\bibitem{Biham} O.\ Biham, D. Mukamel, J. Toner, and X. Zhu, Phys.\ Rev.
Lett.\ \textbf{59}, 2439 (1987).
\end{thebibliography}
\end{document}